\documentclass[doublecol]{epl2}

\title{Temperature dependence of thermal conductivity in 1D nonlinear lattices}

\author{Nianbei Li\inst{1} \and Baowen Li\inst{1,2}}

\institute{
  \inst{1} Department of Physics and Centre for
Computational Science and Engineering, National University of
Singapore, Singapore 117542,
 Republic of Singapore\\
  \inst{2} NUS Graduate School for Integrative Sciences and Engineering,
Singapore 117597, Republic of Singapore
}

\date{28 Feb 2007}
\abstract{ We examine the temperature dependence of thermal
conductivity of one dimensional nonlinear (anharmonic) lattices
with and without on-site potential. It is found from computer
simulation that the heat conductivity depends on temperature via
the strength of nonlinearity. Based on this correlation, we make a
conjecture in the effective phonon theory that the mean-free-path
of the effective phonon  is inversely proportional to the strength
of nonlinearity. We demonstrate analytically and numerically that
the temperature behavior of the heat conductivity
$\kappa\propto1/T$ is not universal for 1D harmonic lattices with
a small nonlinear perturbation. The computer simulations of
temperature dependence of heat conductivity in general 1D
nonlinear lattices are in good agreements with our theoretic
predictions. Possible experimental test is discussed. }

\begin{document}

\maketitle

The role of nonlinearity in the dynamics of one dimensional (1D)
nonlinear (anharmonic) lattices has attracted attention for many
decades, for instance, the ergodicity problem in the
Fermi-Pasta-Ulam (FPU) chains inducted by nonlinear
interactions\cite{FPUT,Ford}, solitons in nonlinear partial
differential equations\cite{KZ} and discrete breathers in
nonlinear lattices\cite{Flach,Flach2}, to name just a few. Recent
years have witnessed increasing studies on the role of
nonlinearity in heat conduction in low dimensional systems (See
Ref.\cite{Review} and the references therein). The fundamental
question is whether the non-integrability or chaos is an essential
or a sufficient condition for the heat conduction to obey the
Fourier law. From computer simulations, it is found that in some
1D nonlinear lattices such as the Frenkel-Kontorova (FK) model and
the $\phi^4$ model, the heat conductivity is size
independent\cite{HLZ98,HLZ00,Aoki1}. Thus the heat conduction in
these models obeys the Fourier's law. This transport behavior is
called normal heat conduction. Whereas in some other nonlinear
lattices such as the FPU and alike models, the heat conduction
exhibits anomalous behavior\cite{Lepri2}, namely the heat
conductivity $\kappa$ diverges with the system size $N$ as $\kappa
\sim N^{\delta}$. Studies in past years have focused on the
physical origin and the value of the divergent exponent
$\delta$\cite{Lepri2,Lepri3,Pereverzev,Narayan,WangLi}. It is
found that the anomalous heat conduction is due to the anomalous
diffusion and a quantitative connection between them has been
established\cite{Li}. Most recently, a very general effective
phonon theory has been proposed to describe the normal and
anomalous heat conduction under the same framework\cite{LTL}.

As a by-product of the study of heat conduction in low dimensional
systems, nonlinearity (anharmonicity) has been found very useful
in controlling heat flow. Because of the nonlinearity, the lattice
vibrational spectrum depends on temperature. This property has
been used to design the thermal
rectifiers/diodes\cite{TPC,LWC,LLW,Hu06,LanLi,Hu1} and thermal
transistors\cite{LWC2}. Inspired by the two segment theoretical
models\cite{LWC,LLW}, Chang et al \cite{Berkely} has built the
first solid state thermal rectifier with a single walled carbon
nanotube and boron nitride nanotube, which indicates the opening
of a new research field - controlling heat flow at microscopic
level by using nonlinearity.

However, in contrast to the size-dependence, the temperature
dependence of heat conductivity and its relationship with the
nonlinearity have not yet been studied systematically even though
this problem is very fundamental and very important from the
experimental point of view. In fact, for low dimensional nanoscale
systems such as nanotube and nanowires etc, to measure the
temperature dependence of thermal conductivity for a fixed length
sample is much easier than that for the size dependence of thermal
conductivity for a fixed temperature. The difficulty for measuring
size-dependent thermal conductivities lies in the fabrication of
low dimensional systems. One cannot guarantee that the systems
(nanotube/nanowires) of different system sizes fabricated
separately are identical. However, performing the experiments at
different temperatures for a fixed length is feasible.

In this Letter, we study the temperature dependence of heat
conductivity and its connection with nonlinearity in several
representative 1D nonlinear lattices with and without on-site
potential, namely, the FK model, the $\phi^4$ model, and the FPU
models. We will show that the temperature dependence of heat
conductivity cannot be understood by the phenomenological phonon
collision theory\cite{Kittel} in harmonic lattice with a small
nonlinear perturbation. We find via computer simulations that the
heat conductivity is actually inversely proportional to the
strength of nonlinearity. More interestingly, if we make a
conjecture in the effective phonon theory that the  mean-free-path
of effective phonon is inversely proportional to the strength of
nonlinearity, we can give a rather satisfactory explanation for
the temperature dependence of heat conductivities at both weak and
strong coupling regimes in 1D nonlinear lattices such as the FK
and the FPU-$\beta$ models {\it consistently}. We argue that the
results presented in the current paper can be tested by nano scale
experiments.

Without loss of generality and simplicity, we consider a 1D
nonlinear lattice whose Hamiltonian reads,
\begin{equation}
H=\sum^{N}_{i=1}\left[\frac{p^2_i}{2}+V(x_i,x_{i+1})+U(x_i)\right],
\end{equation}
where $x_i$ is the displacement from the equilibrium position of
the $i\it{th}$ particle. In the following study we consider both
cases with on-site potential, such as the FK model and the
$\phi^4$ model, and the cases without on-site potential such as
the FPU-$\alpha$, the FPU-$\beta$ model and the FPU-$\alpha\beta$
model.

{\it The Frenkel-Kontorova model} In the FK model
$V=\frac{1}{2}(x_i-x_{i+1})^2$, and
$U=\frac{K}{(2\pi)^2}(1-\cos{2\pi x_i})$.  In this Letter, we take
parameter $K=2\pi$. The FK model is very special since it reduces
to a harmonic lattice with a small nonlinear perturbation at both
low and high temperature regimes. According to the phonon
collision theory, the collision frequency of a given phonon should
be proportional to the number of phonons with which it can
collide. Since the phonon distribution function is proportional to
$T$ in classical regime, the total number of excited phonons is
proportional to $T$. Hence the mean-free-path of  phonon
$l\propto1/T$. At this regime, the heat capacity and velocity of
phonons are all constants. From the Debye formula, one gets the
heat conductivity, $\kappa\propto l$. Therefore the heat
conductivity, $\kappa\propto 1/T$ universally for the harmonic
lattice with a small nonlinear perturbation.

\begin{figure}
\includegraphics[width=\columnwidth]{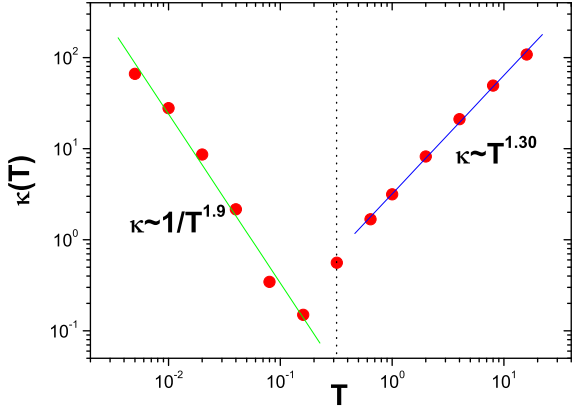}
\vspace{-1.5cm}\\
 \caption{(Color online)Heat conductivity $\kappa$ vs temperature, $T$,
for the FK model with lattice length $N=500$. The best fit line is
$\kappa\propto T^{-1.9\pm 0.1}$ at the low temperature regime and
is $\kappa\propto T^{1.30\pm 0.03}$ at the high temperature
regime. The dotted line in the middle is the turning point
estimated from the Hamiltonian, $T_0=2K/(2\pi)^2$, which is
$1/\pi$, in the case of $K=2\pi$ in this paper.} \label{fig:FK}
\end{figure}

We compute the heat conductivity of the FK model as a function of
temperature with fixed boundary conditions and the Nose-Hoover
heat baths. The results are shown in Fig.\ref{fig:FK} in double
logarithmic scales. At the high temperature regime, the on-site
potential is considered as a small perturbation, thus the FK model
is suitable for the interpretation of phonon collision theory.
However the heat conductivity in this regime is not inversely
proportional to temperature $\kappa\propto 1/T$ as the phonon
collision theory predicts, it increases almost linearly with the
temperature $\kappa\propto T$! This striking contradictory
suggests that the phonon collision theory may not be suitable to
describe the classical heat transport.

To explain the temperature dependence of heat conductivity in 1D
nonlinear lattices, we introduce a new quantity to describe the
strength of nonlinearity. We define a dimensionless
 nonlinearity $\epsilon$ as a ratio between the average
of nonlinear potential energy and the total potential energy which
consists of linear and nonlinear potential energy:
\begin{equation}
\epsilon=\frac{\left|\langle E_n\rangle\right|}{\langle
E_l+E_n\rangle},\,\,0\leq\epsilon\leq1 ,
\end{equation}
where $\langle\cdot\rangle$ is an ensemble average. For the
harmonic lattice with a small nonlinear perturbation, the average
of nonlinear potential energy $\langle E_n\rangle$ is negligible
compared to the average of linear potential energy $\langle
E_l\rangle$, thus the dimensionless nonlinearity can be
approximated by $\epsilon\approx\frac{\left|\langle
E_n\rangle\right|}{\langle E_l\rangle}$, where the average of the
linear potential energy $\langle E_l\rangle=Nk_{B}T/2$ at the
harmonic limit according to the theory of energy
equipartition\cite{Huang}.

At the high temperature regime, the average of nonlinear on-site
potential energy $\left<E_n\right>$ is negligible compared to the
average of harmonic inter particle potential energy
$\left<E_l\right>$. The dimensionless nonlinearity can be
expressed as
\begin{equation}
\epsilon \approx
\frac{\left<\sum_{i}\frac{1}{2\pi}\left(1-\cos{2\pi
x_i}\right)\right>}
{\left<\sum_{i}\frac{1}{2}\left(x_i-x_{i+1}\right)^2\right>}
\propto
\frac{N}{\left<\sum_{i}\frac{1}{2}\left(x_i-x_{i+1}\right)^2\right>},
\end{equation}
where the average of the on-site potential energy of any particle,
$\left<1-\cos{2\pi x_i}\right>$, has been assumed to be a constant
in the interval $(0,2)$ at the high temperature regime. The
average of inter particle potential energy
$\left<\sum^{N}_{i=1}\frac{1}{2}\left(x_i-x_{i+1}\right)^2\right>=Nk_{B}T/2$.
Therefore, the dimensionless nonlinearity at high temperature
regime has the asymptotic behavior: $\epsilon\propto 1/T$.
Compared with the computer simulation it is not difficult to find
the correlation between the heat conductivity and the
dimensionless nonlinearity at the high temperature regime for the
FK model:
\begin{equation}
\kappa\propto \frac{1}{\epsilon}\propto T.
\end{equation}
This correlation strongly suggests that the mean-free-path of
phonons should be proportional to the inverse of the dimensionless
 nonlinearity, $l\propto 1/\epsilon$, rather than to
the inverse of temperature, $l\propto 1/T$, as the phonon
collision theory predicts.

As for the nonlinear lattice, the notion of phonon is not valid
any more, one should invoke the effective phonon theory\cite{LTL}.
Recently, we find out that the normal and anomalous heat
conduction can be treated with effective phonon theory\cite{LTL}.
For a general 1D nonlinear lattice with
\begin{equation}
V(x_i,x_{i+1})=\sum^{\infty}_{s=2}g_{s}\frac{\left(x_i-x_{i+1}\right)^{s}}{s},
U(x_i)=\sum^{\infty}_{s=2}\sigma_{s}\frac{x^s_i}{s},
\end{equation}
the spectrum of effective phonons is
$\hat{\omega}^2_k=\alpha(\omega^2_k+\gamma)$, where
$\omega=2\sqrt{g_2}\sin{k/2}$ is the phonon dispersion of harmonic
lattice, and the temperature dependent coefficients $\alpha$ and
$\gamma$ are defined as
\begin{equation}
\alpha=\frac{\sum_{s}g_{s}\langle\sum_{i}(x_i-x_{i+1})^s\rangle}
{\langle\sum_{i}(x_i-x_{i+1})^2\rangle},
\gamma=\frac{1}{\alpha}\frac{\sum_{s}\sigma_{s}\langle\sum_{i}x^s_i\rangle}
{\langle\sum_{i}x^2_i\rangle}.
\end{equation}
The heat conductivity can be expressed by the modified Debye
formula:
\begin{equation}
\kappa=\frac{c}{2\pi}\int^{2\pi}_{0}P(k)v(k)l(k)dk,
\end{equation}
where $c$ is the specific heat, $P(k)$ the normalized power
spectrum of total heat flux,
$v(k)=\partial{\hat{\omega}_k}/\partial{k}$ the velocity of the
effective phonon, and $l(k)$ the mean-free-path of effective
phonon. $l(k)=v(k)\tau(k)$, where $\tau(k)$ is the effective
phonon relaxation time which is proportional to the quasi-period
of effective phonons, $\tau(k)=\lambda 2\pi/\hat{\omega}_k$, and
the dimensionless $\lambda$ only depends on temperature.

Here we conjecture that the  mean-free-path of the effective
phonon is inversely proportional to the dimensionless
nonlinearity,
\begin{equation}
\lambda\propto \frac{1}{\epsilon}.
\end{equation}
In the harmonic limit $\epsilon\rightarrow 0$, the phonons should
have infinite mean-free-path which is exactly implied by this
conjecture. The temperature dependence of heat conductivity can be
expressed as
\begin{equation}
\kappa= \frac{\lambda c\sqrt{\alpha}}{2\pi} {\cal P} \propto
\frac{c\sqrt{\alpha}}{\epsilon} {\cal P},
\end{equation}
where
$$
 {\cal P}=
\int^{2\pi}_{0}P(k)\frac{\omega^2_k}{(\omega^2_k+\gamma)^{3/2}}\cos^2{\frac{k}{2}}dk.$$
 The low and high temperature regimes are of special interest, in
which the heat conductivity exhibits asymptotic behaviors. The
transport properties of general nonlinear lattices are mostly
determined by the leading terms of potential energy perturbed by
another potential energy with the first order approximation. For
example, the dimensionless FPU-$\beta$ lattice has quadratic and
quartic inter particle potential energy
\begin{equation}
H=\sum^{N}_{i=1}\left[\frac{p^2_{i}}{2}+\frac{1}{2}(x_i-x_{i+1})^2
+\frac{1}{4}(x_i-x_{i+1})^4\right].
\end{equation}
At low temperature regime, the leading term of potential energy is
the quadratic, and the quartic term is regarded as a small
perturbation. Whereas at high temperature regime, the leading term
is the quartic one, and the quadratic term is a small
perturbation. At these temperature regimes we can assume that
$P(k)$ is temperature independent since the topology of the
Hamiltonian doesn't change with temperature. Furthermore, the
unchanged topology of the Hamiltonian also implies that the
specific heat $c$ is almost temperature independent. In the
classical case, $c=\langle H\rangle/NT$ where $\langle H\rangle$
is the ensemble average of total energy at temperature $T$. The
specific heat $c=1$ at classic harmonic lattice if we take the
Boltzman constant $k_{B}$ to be unity. We take the FPU-$\beta$
lattice as example. At low temperature regime the FPU-$\beta$
lattice reduces to a harmonic lattice, so $c\approx1$. At the high
temperature regime, the leading order of potential energy is the
quartic potential energy. Due to the general equipartition theory,
$NT=\left<\sum_{i}x_{i}\frac{\partial{H}}{\partial{x_{i}}}\right>\approx\left<\sum_{i}(x_i-x_{i+1})^4\right>$.
Then the specific heat
$c\approx\left<\sum_{i}p^2_i/2\right>/NT+\left<1/4\sum_{i}(x_i-x_{i+1})^4\right>/NT=3/4$
where we have ignored the harmonic potential at high temperature
regime. The specific heat $c$ is just a constant in both
temperature regions. This is true for general 1D nonlinear
lattices, thus we can write,
\begin{equation}\label{kappa}
\kappa\propto \frac{\sqrt{\alpha}}{\epsilon} {\cal P}.
\end{equation}
Therefore, the heat conductivity depends on temperature via the
dimensionless nonlinearity $\epsilon$. The two coefficients
$\alpha$ and $\gamma$ can be calculated from statistical
mechanics.

For the FK model, the coefficient $\alpha=1$ exactly. At the high
temperature regime, the dimensionless nonlinearity
$\epsilon\propto 1/T$ as we have discussed above, thus the heat
conductivity depends on temperature as
\begin{equation}
\kappa\propto T {\cal P}.
\end{equation}
Since $\gamma$ goes to $0$ as the temperature goes to infinity,
$\gamma$ should decrease with the increase of temperature. As a
result, ${\cal P}$ should increase with temperature $T$.
Therefore, the heat conductivity should increase faster than a
linear dependence of $T$. This is what we observe in numerical
results shown in Fig.\ref{fig:FK}, where $\kappa\propto T^{1.30}$
at the high temperature regime.

At the low temperature regime, the on-site potential can be
expanded in a Taylor series:
\begin{equation}
U(x_i)=\frac{1}{2\pi}\left[\frac{(2\pi x_i)^2}{2!}-\frac{(2\pi
x_i)^4 }{4!}+\frac{(2\pi x_i)^6 }{6!}-...\right].
\end{equation}
The linear potential energy consists of the harmonic inter
particle potential energy and the harmonic on-site potential
energy,
$E_l=\sum^{N}_{i=1}\left[\frac{1}{2}(x_i-x_{i+1})^2+2\pi\frac{
x_i^2}{2!}\right]$. The equipartition of energy still holds,
$\langle E_l\rangle=Nk_{B}T/2$. The first order nonlinearity is
estimated by the average of quartic on-site potential over the
linear potential energy, $\epsilon\approx
\langle\sum^{N}_{i=1}\frac{1}{2\pi}\frac{(2\pi x_i)^4
}{4!}\rangle/\langle E_l\rangle\propto T$. However, the particles
are not stable in the Hamiltonian with this first order nonlinear
approximation because of the minus sign before the quartic on-site
potential. At this low temperature regime, the system is most
likely to approach a stable state with the second order nonlinear
approximation. The proper estimation of the dimensionless
nonlinearity should be proportional to
$\langle\sum^{N}_{i=1}\frac{1}{2\pi}\frac{(2\pi
x_i)^6}{6!}\rangle/\langle E_l\rangle\propto T^2$. Therefore, the
heat conductivity should depend on temperature as $\kappa\propto
1/T^2$ if we neglect the slight temperature dependence caused by
the coefficient $\gamma$. This is also in a good agreement with
the numerical result $\kappa\propto 1/T^{1.9}$ in
Fig.\ref{fig:FK}. The slight difference might come from the first
order unstable nonlinear potential and the temperature dependence
caused by the coefficient $\gamma$ which has been neglected in
this situation.

From Fig.1 it is clear that the heat conductivity decreases at low
temperature region and increases at high temperature region. The
turning point can be roughly estimated from the Hamiltonian
itself. Here we use the expression of the specific heat again
\begin{equation}
c=\frac{\left<\sum_{i}\left[\frac{p^2_{i}}{2}+\frac{1}{2}(x_i-x_{i+1})^2+\frac{K}{(2\pi)^2}(1-\cos{2\pi
x_i})\right]\right>}{NT}
\end{equation}
By definition, $\langle\sum_{i}p^2_i/2\rangle=NT/2$. At this
turning point, we expect the ensemble average of inter particle
potential and the on-site potential to be the same value
\begin{equation}
\langle\sum_{i}\frac{1}{2}(x_i-x_{i+1})^2\rangle=\langle\sum_{i}\frac{K}{(2\pi)^2}(1-
\cos{2\pi x_i})\rangle.
\end{equation}
The temperature of the turning point thus can be expressed as
\begin{equation}
T_0=\frac{\frac{2K}{(2\pi)^2}}{c-1/2}\frac{\langle\sum_{i}(1-\cos{2\pi
x_i})\rangle}{N} \approx\frac{2K}{(2\pi)^2},
\end{equation}
where we have used the approximations $c\approx1$ and
$\left<\sum_{i}(1-\cos{2\pi x_i})\right>\approx N/2$. With
$K=2\pi$ in our numerical calculation, we have $T_0=1/\pi$, which
is shown as the dotted line in the middle of Fig.1. It is in a
good agreement with the numerical calculations.

{\it The $\phi^4$ model} Another commonly studied model is the
$\phi^4$ model which has an  on-site potential
$U(x)=\frac{1}{4}x^4$. The coefficient $\alpha=1$. At the low
temperature regime, the $\phi^4$ model is topologically similar to
the FK model. The nonlinear quartic on-site potential in the
$\phi^4$ model guarantees the particles in stable states, thus the
heat conductivity depends on temperature as $\kappa\propto {\cal
P}/T$. ${\cal P}$ decreases with temperature because $\gamma$
increases with temperature. Thus the heat conductivity should
decrease faster than $\kappa\propto 1/T$ which is consistent with
the numerical results in Ref.\cite{Aoki1} where $\kappa\propto
1/T^{1.35}$. The high temperature regime of the $\phi^4$ model is
not in our consideration because it has no energy transport at
high temperature limit which is also called anti-continuous or
anti-integrable limit\cite{Flach}.

{\it The FPU-$\beta$ model} In the FPU-$\beta$ model, $U(x)=0$,
and $V(x)=
\frac{k}{2}(x_i-x_{i+1})^2+\frac{\beta}{4}(x_i-x_{i+1})^4$. We
take $k=1$ and $\beta=1$. For the general 1D nonlinear lattices
without on-site potential such as the FPU-$\beta$ model, they have
total different dynamical properties at low temperature (weak
coupling) regime and high temperature (strong coupling) regime.
The leading order potential energy is quadratic potential at low
temperature regime and quartic potential at high temperature
regime. The coefficient $\gamma=0$ for general 1D nonlinear
lattices without  on-site potential. This property allows us  to
predict the  dependence of heat conductivity more precisely since
${\cal P}$ is a constant. In this case,
\begin{equation}
\kappa\propto \frac{\sqrt{\alpha}}{\epsilon}.
\end{equation}

At low temperature regime, the coefficient $\alpha\approx 1$ by
definition and the heat conductivity is just proportional to
$1/\epsilon$. The dimensionless nonlinearity for the FPU-$\beta$
model is $\epsilon\approx \langle
\sum_{i}\frac{1}{4}(x_i-x_{i+1})^4\rangle/\left<\sum_{i}\frac{1}{2}(x_i-x_{i+1})^2\right>\propto
T$. Thus the heat conductivity $\kappa \propto 1/T$ which is
consistent with the numerical results in Ref.\cite{Aoki2} and in
the Fig.2.

\begin{figure}
\includegraphics[width=\columnwidth]{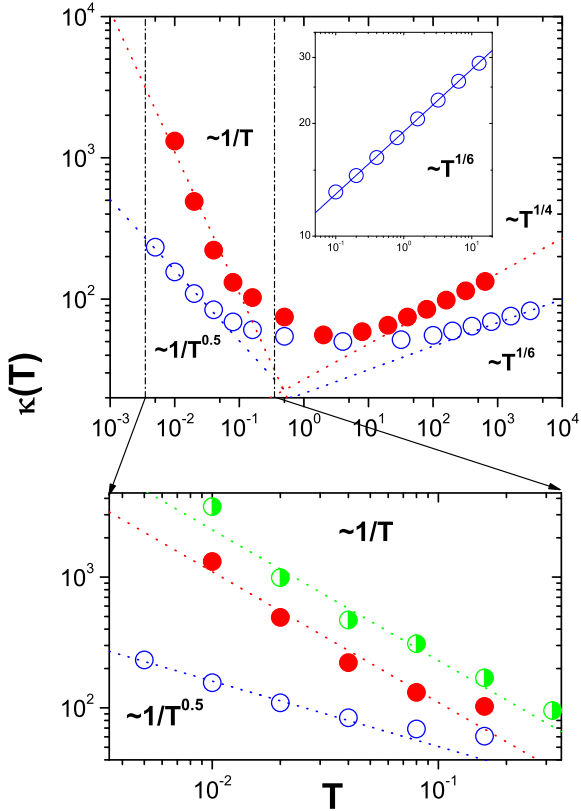}
\vspace{-.5cm}
 \caption{(Color-online)(a) Heat conductivity $\kappa$ vs  $T$
for the FPU lattices with $N=2000$. The solid circles are
numerical results of the FPU-$\beta$ model. Open circles are that
of the FPU-$\alpha$ model with absolute cubic potential where data
have been shifted vertically for comparison. The inset is high
temperature behavior of the FPU-$\alpha$ model which is actually
calculated with parameters $k=0$ and $\alpha^*=1$. $k=0$ is
equivalent to the high temperature limit. The lower panel is the
enlargement of low temperature part. The additional semi circles
are from the FPU-$\alpha\beta$ model with cubic and quartic
interparticle potential.} \label{fig:kappa-T}
\end{figure}

At high temperature regime for the FPU-$\beta$ model, the linear
potential energy is negligible compared to the nonlinear quartic
potential energy. The dimensionless nonlinearity is
$\epsilon=\frac{\left<\sum^{N}_{i=1}\frac{1}{4}(x_i-x_{i+1})^4\right>}
{\left<\sum^{N}_{i=1}\frac{1}{2}(x_i-x_{i+1})^2+\frac{1}{4}(x_i-x_{i+1})^4\right>}
\approx 1$. The heat conductivity only depends on coefficient
$\alpha$ at this region as
\begin{equation}
\kappa \propto \sqrt{\alpha}.
\end{equation}
The coefficient $\alpha$ can be exactly calculated from the formula in Ref.\cite{LTL}
\begin{eqnarray}
\alpha&=&1+\frac{\int^{\infty}_{0}\phi^{4}e^{-\left(\frac{\phi^2}{2}+\frac{\phi^4}{4}\right)/T}d\phi}
{\int^{\infty}_{0}\phi^{2}e^{-\left(\frac{\phi^2}{2}+\frac{\phi^4}{4}\right)/T}d\phi}\nonumber\\
&\approx&\frac{\int^{\infty}_{0}\phi^{4}e^{-\frac{\phi^4}{4T}}d\phi}
{\int^{\infty}_{0}\phi^{2}e^{-\frac{\phi^4}{4T}}d\phi}
=2\frac{\Gamma(\frac{5}{4})}{\Gamma(\frac{3}{4})}T^{1/2},
T\rightarrow\infty.
\end{eqnarray}
Therefore, the heat conductivity at high temperature regime is
$\kappa\propto T^{1/4}$. This temperature behavior is consistent
with previous results in Ref.\cite{Aoki2} and our numerical
results in Fig.\ref{fig:kappa-T}.

 {\it The FPU-$\alpha$ model}
We consider the FPU-$\alpha$ model in which
$V=\frac{k}{2}(x_i-x_{i+1})^2+\frac{\alpha^*}{3}\left|x_i-x_{i+1}\right|^3$.
We take $k=1$ and $\alpha^*=1$. Here the absolute cubic inter
particle potential is used to prevent the particles from escaping
to infinity which occurs in the usual FPU-$\alpha$ model. At the
low temperature regime of the FK model, we have met the case that
the heat conductivity doesn't decrease with temperature as
$\kappa\propto 1/T$ which is believed to be universal in solid
state physics. Here we present another case that the heat
conductivity is not proportional to $1/T$ at the low temperature
regime.

Fig.\ref{fig:kappa-T} shows the numerical results of heat
conductivity of this modified FPU-$\alpha$ model as a function of
temperature. At the low temperature regime, $\kappa\propto
1/\sqrt{T}$. The dimensionless nonlinearity $\epsilon\approx
\left<\sum_{i}\frac{1}{3}\left|x_i-x_{i+1}\right|^3\right>/\left<\sum_{i}\frac{1}{2}(x_i-x_{i+1})^2\right>\propto
\sqrt{T}$. Thus the prediction derived from our conjecture
$\kappa\propto 1/\epsilon\propto 1/\sqrt{T}$ at low temperature
regime is in good agreement with the numerical results.

At high temperature regime, the coefficient $\alpha$ is
\begin{eqnarray}
\alpha&=&1+\frac{\int^{\infty}_{0}\phi^{3}e^{-\left(\frac{\phi^2}{2}+\frac{\phi^3}{3}\right)/T}d\phi}
{\int^{\infty}_{0}\phi^{2}e^{-\left(\frac{\phi^2}{2}+\frac{\phi^3}{3}\right)/T}d\phi}\nonumber\\
&\approx&\frac{\int^{\infty}_{0}\phi^{3}e^{-\frac{\phi^3}{3T}}d\phi}
{\int^{\infty}_{0}\phi^{2}e^{-\frac{\phi^3}{3T}}d\phi}=3^{1/3}\Gamma\left(\frac{4}{3}\right)T^{1/3}.
\end{eqnarray}
The heat conductivity should depend on temperature as
$\kappa\propto\sqrt{\alpha}\propto T^{1/6}$. The numerical results
in Fig.\ref{fig:kappa-T} confirm this prediction. For the general
realistic solids, the inter particle potential can always be
expanded as a polynomial form. However, the first order nonlinear
potential is cubic where particles will escape to infinity.
Particles are more likely to approach the stable states governed
by the next order nonlinear potential: quartic potential. This
might be the reason that the realistic solids have the universal
behavior ($\kappa\propto 1/T$) experimentally at this regime. To
check this in 1D we calculate the temperature dependence of heat
conductivity of the FPU-$\alpha\beta$ model which has both cubic
and quartic inter particle potential besides the harmonic inter
particle potential, namely,
$V(x_i,x_{i+1})=\frac{k}{2}(x_i-x_{i+1})^2+\frac{\alpha^*}{3}(x_i-x_{i+1})^3
+\frac{\beta}{4}(x_i-x_{i+1})^4$ (with $k=1$, $\alpha^*=1$,
$\beta=1$)
 and find the temperature dependence of $\kappa$ is more
like that of FPU-$\beta$ model (lower panel of Fig.\ref{fig:kappa-T}).

Finally we discuss the possible experiment to verify our
conjecture. From the computer simulations we can see that
eventually the nonlinearity will cause the heat conductivity to
bend upward as the temperature increases. It is worthwhile to
point out that the $\phi^4$ model does not show this property
because it cannot model realistic solids in the high temperature
regime. It is not physical to make the substrate on-site potential
larger than the interparticle potential which binds the atoms
together. The reason we cannot observe this kind of temperature
dependence of heat conductivity in realistic solids at room
temperature is that the strength of nonlinearity in them is too
weak. The realistic solids just start melting before the
nonlinearity dominates the transport. However, with nano
technology we can build the 1D sample attached to a substrate to
demonstrate this counter-intuitive temperature behavior of heat
conductivity at room temperature caused by nonlinearity. Now we
consider the proposed lattice model with real physical units:
\begin{equation}
H=\sum^{N}_{i=1}\left[\frac{p^{2}_{i}}{2m}+\frac{1}{2}m\omega^2(x_{i+1}-x_{i})^2+U(x_i)\right]
\end{equation}
where $m$ is the mass of atom, $\omega$ is the oscillation
frequency, and $U(x_i)$  the weak substrate on-site potential with
strength $A$ which can be controlled by experiment. As we have
discussed above, this on-site potential $U(x_i)$ does not need to
be the FK like potential. $U(x_i)$ can be any kind of on-site
potential with strong nonlinearity. Here the ``strong" means the
linear part inside $U(x_i)$ is small compared with the nonlinear
part. If the magnitude of this on-site coupling strength is $A$,
then using the same analysis we have used for the FK model, the
heat conductivity should increase with temperature after
\begin{equation}
T_0=\frac{2A}{k_{B}}\,\,\,\, (Kelvin),
\end{equation}
and we predict the increase is larger than the linear increase
with temperature. To observe this classical behavior explicitly,
we must lie outside of the quantum regime where the increase of
heat conductivity is mostly caused by the increase of specific
heat with temperature. Finally, we must emphasize that the key
point for this experiment is the substrate on-site potential with
``strong" nonlinearity.

In summary, we have found via computer simulation  the strong
correlation between the temperature dependence of heat
conductivity and the strength of nonlinearity in 1D nonlinear
lattices. In the effective phonon theory of heat conduction, we
conjecture that the  mean-free-path of effective phonon is
inversely proportional to the strength of dimensionless
nonlinearity. The predictions from this conjecture are consistent
with the numerical results in several very general 1D nonlinear
lattices with and without on-site potential. We found the
temperature dependence of heat conductivity $\kappa\propto 1/T$ is
not universal for the model of harmonic lattice with a small
nonlinear perturbation. The temperature dependence of heat
conductivities in the commonly studied FK and FPU models are
explained by our theory at both weak and strong coupling regimes
{\it consistently}. We have also proposed possible experiment to
verify our conjecture.

\acknowledgments
This work is supported in part by a FRG grant of NUS and the DSTA
under Project Agreement POD0410553.

\end{document}